\documentclass[aps,reprint,twocolumns,pre,superscriptaddress,floatfix,notitlepage,nofootinbib]{revtex4-1}
\usepackage{amssymb,amsmath,amsfonts} 
\usepackage{mathtools}
\usepackage{physics}
\usepackage{array}
\usepackage{graphicx,epsfig,xcolor}
\usepackage[colorlinks=true,citecolor=blue,linkcolor=red,pdfauthor={Corps and Relano}]{hyperref}
\usepackage{newtxtext,newtxmath}

\newcommand{\hatmath}[1]{\hat{\mathcal{#1}}} 

\begin{document}

\title{Scarred ferromagnetic phase in the long-range transverse-field Ising model}

\author{\'{A}ngel L. Corps}
    \email[]{corps.angel.l@gmail.com}
    \affiliation{Institute of Particle and Nuclear Physics, Faculty of Mathematics and Physics, Charles University, V Hole\v{s}ovi\v{c}k\'{a}ch 2, 180 00 Prague, Czech Republic}
    \affiliation{Grupo Interdisciplinar de Sistemas Complejos (GISC),
Universidad Complutense de Madrid, Avenida Complutense s/n, E-28040 Madrid, Spain}
    
\author{Armando Rela\~{n}o}
    \email[]{armando.relano@fis.ucm.es}
        \affiliation{Grupo Interdisciplinar de Sistemas Complejos (GISC),
Universidad Complutense de Madrid, Avenida Complutense s/n, E-28040 Madrid, Spain}
\affiliation{Departamento de Estructura de la Materia, F\'{i}sica T\'{e}rmica y Electr\'{o}nica, Universidad Complutense de Madrid, Avenida Complutense s/n, E-28040 Madrid, Spain}

\date{\today} 

\begin{abstract}

  We report the existence of a large set of ferromagnetic scarred states in the one-dimensional transverse-field Ising model with long-range interactions, in a regime with no ferromagnetic phase at finite temperature. These scarred states are distributed over different spectral regions, surrounded by a paramagnetic sea. We show that simple initial conditions, consisting in a few small magnetic domains, selectively populate these scarred states. This leads to the appearance of a special dynamical phase, which we call scarred ferromagnetic phase. As a consequence, initial states with a small number of small magnetic domains evolve towards ferromagnetic equilibrium states, whereas initial states with larger domains or no magnetic structure relax to the expected thermal paramagnetic equilibrium state. 
\end{abstract}

\maketitle

\textit{Introduction.--} Many-body quantum systems with long-range interactions play an increasingly relevant role in quantum physics \cite{Defenu2023,Defenu2024}. These interactions induce a number of exotic physical phenomena. For example, they allow for the existence of ordered phases at finite temperature in one-dimensional systems \cite{Dyson1969,Thouless1969,Fisher1972,Liujten1997,Gonzalez-Lazo2021,Corps2024pre,Schuckert2025}, although these phases are strictly forbidden with standard short-range interations \cite{Landau1980}. They are also linked to time crystals \cite{Zhang2017,Choi2017,Collura2022,Gargiulo2024,Pizzi2021,Yu2019,Yu2025}, ensemble inequivalence \cite{Russomanno2021}, prethermalization and the emergence of long-lived stationary states \cite{Halimeh2017b,Neyenhuis2017,Defenu2021,Mattes2025}, dynamical quantum phase transitions \cite{Zhang2017b,Halimeh2017,Homrighausen2017,Zunkovic2018,Piccitto2019,Ranabhat2022,Corps2022,Corps2022prl}, quantum hydrodynamics \cite{Joshi2022}, condensed-matter analogues of quark confinement \cite{Hauke2013,Lerose2019,Liu2019,Tan2021} and dynamical hadron formation \cite{Verdel2020,Vovrosh2022}. Long-range interacting systems are experimentally accessible, with a controllable interaction decay, on several experimental platforms, such as cavity-atom systems \cite{Mivehvar2021}, trapped ions \cite{Britton2012,Feng2023} or Rydberg atoms \cite{Labuhn2016,Zeiher2017,Chen2023}, and they can also be efficently simulated \cite{Achutha2025}. Unveiling the properties of these systems is crucial for advancing quantum simulation and for probing fundamental aspects of quantum many-body physics far from equilibrium.

In this Letter, we report the existence of a set of states, which we call \textit{ferromagnetic scars}, in the absence of a ferromagnetic phase at finite temperature. If an initial state is chosen such that they are populated, the ensuing equilibrium state is also ferromagnetic, although the thermal equilibrium with the same energy is paramagnetic. Therefore, these ferromagnetic scars play a role very similar to that of quantum many-body scars in chaotic systems \cite{Turner2018,Lerose2025}: they violate ergodicity, giving rise to ordered symmetry-breaking states in regions in which every initial condition is expected to evolve into a disordered symmetric equilibrium state. Focusing on the one-dimensional transverse field Ising model, we show that the ferromagnetic scars are selectively populated by initial states consisting of a small number of magnetic domains of small relative size. Then, we propose that this model displays a dynamical phase transition induced by the number and the size of magnetic domains in the initial condition. 

\textit{Model.--} We consider the one-dimensional transverse-field Ising model (TFIM) with power-law interactions,
\begin{equation}\label{eq:TFIM}
    \hat{H}_{\textrm{TFIM}}=-\frac{1}{\mathcal{K}(\alpha)}\sum_{i<j} \frac{J}{|i-j|^{\alpha}}\hat{\sigma}_{i}^{x}\hat{\sigma}_{j}^{x}+h\sum_{i}\hat{\sigma}_{i}^{z}.
\end{equation}
Here, $J$ is a coupling constant, $h$ is a transverse magnetic field along the z-axis, and $\alpha$ accounts for the power-law interaction between spins. $\mathcal{K}(\alpha)$ is the so-called Kac factor \cite{Kac1963}, which is necessary to ensure that the Hamiltonian is extensive for $\alpha<1$. For details on its computation, see Appendix \ref{sec:simulationTFIM}. The case $\alpha=0$ is the fully-connected limit, where Eq. \eqref{eq:TFIM} coincides with the Lipkin-Meshkov-Glick model \cite{Lipkin1965}, whose behavior has been studied extensively, including quantum phase transitions, excited-state quantum phase transitions, and dynamical phase transitions \cite{Corps2022,Corps2022prl,Botet1982,Homrighausen2017,Cejnar2021,Puebla2015,Puebla2013b}. For $\alpha\to\infty$, we recover the prototypical nearest-neighbors quantum Ising model \cite{Sachdev}.

The Hamiltonian Eq. \eqref{eq:TFIM} is invariant under the parity operator $\hat{\Pi}=\prod_{i=1}^{N}\hat{\sigma}_{i}^{z}$, which allows us to classify its eigenstates according to positive and negative values of this $\mathbb{Z}_{2}$ operator, $\hat{\Pi}\ket{E_{n,\pm}}=\pm \ket{E_{n,\pm}}$. It has a ferromagnetic phase in which this $\mathbb{Z}_2$ symmetry is broken for $h<h_{c}$ and $T<T_{c}$, if $\alpha \leq 2$ \cite{GonzalezLazo2021}. The total magnetization along the x-axis, $\hat{M}=\frac{1}{2}\sum_{i=1}^{N}\hat{\sigma}_{i}^{x}$, is the order parameter of the corresponding phase transition. For larger values of $\alpha$, ferromagnetism only occurs at $T=0$ and $h<h_c$, but particle confinement \cite{Liu2019,Vovrosh2022}, dynamical phase transitions \cite{Homrighausen2017,Corps2022prl,Sehrawat2021,Halimeh2017,Halimeh2020}, and discrete crystalline time response \cite{Collura2022} have been observed at finite excitation energies.

For our simulations we use periodic boundary conditions, so the Hamiltonian is also invariant under translational transformations as well as inversion with respect to any symmetry axis of the chain; a discussion on these symmetries is given in Appendix \ref{sec:simulationTFIM}. To compare systems of different sizes, we always work with an intensive version of the order parameter, $\hat{m} = 2 \hat{M}/N$.

\begin{center}
\begin{figure}[h!]
\hspace*{-1.1cm}\includegraphics[width=0.54\textwidth]{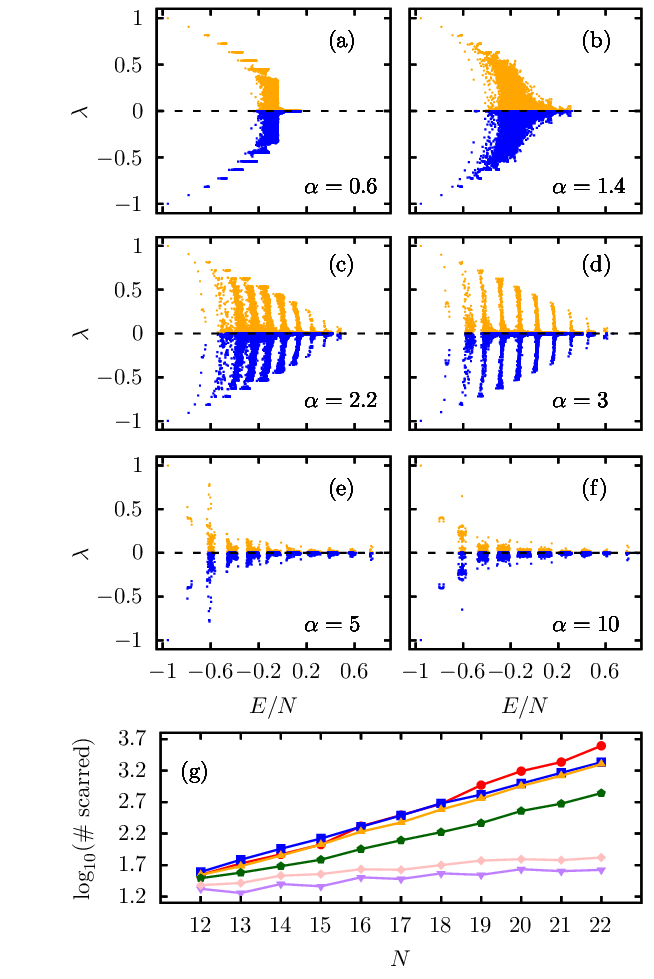}
\caption{(a)-(f) Eigenvalues of the normalized total magnetization in $\mathcal{E}_n$ subspaces for different values of the power-law interaction, $\alpha$, $J=1$ and $h=0.1$, as a function of the energy of the subspace. System size is $N=22$. (g) Scaling of the number of scarred states with absolute values of the eigenvalues of the total magnetization greater than $0.2$ for $\alpha=0.6$ (red), $1.4$ (blue), $2.2$ (orange), $3$ (green), $5$ (purple) and $10$ (pink).}
\label{fig:phases}
\end{figure}
\end{center}

\textit{Thermalization and symmetry-breaking.--} Our understanding of quantum thermalization in closed many-body systems is based on the eigenstate thermalization hypothesis (ETH) \cite{Deutsch1991,Srednicki1994,Rigol2008,Polkovnikov2011,Gogolin2016,DAlessio2016}. According to this theory, the eigenstates of chaotic quantum systems are thermal. As a consequence, the dynamics of generic few-body observables, $\hat{O}$, relax to effective equilibrium states captured by thermal ensembles. Consider an initial state $\ket{\psi_{0}}=\sum_{n}c_{n}\ket{E_{n}}$, where $\{\ket{E_{n}}\}$ denote the eigenstates of a certain Hamiltonian, $\hat{H}\ket{E_{n}}=E_{n}\ket{E_{n}}$. Then, the time-evolved wavefunction reads $\ket{\psi(t)}=\sum_{n}c_{n}e^{-iE_{n}t}\ket{E_{n}}$ ($\hbar=1$), and the equilibrium value of $\hat{O}$ is given by the long-time average $\overline{\hat{O}}=\lim_{\tau\to\infty}\frac{1}{\tau}\int_{0}^{\tau}\textrm{d}t\,\langle \hat{O}(t)\rangle$. For systems without spectral degeneracies, this expression depends exclusively on diagonal eigenstate expectation values of $\hat{O}$, $\overline{\hat{O}}=\sum_{n}|c_{n}|^{2}O_{nn}$, $O_{nn}\equiv \bra{E_{n}}\hat{O}\ket{E_{n}}$; and for chaotic systems fulfilling the ETH, then $\overline{\hat{O}}$ can be described by the microcanonical ensemble, $\langle \hat{O}\rangle_{\textrm{ME}}=\frac{1}{\mathcal{N}}\sum_{E\in[\langle E\rangle-\delta E,\langle E\rangle+\delta E]}O_{nn}$ within a small energy window $\delta E$ around the target energy $\langle E\rangle=\bra{\psi(t)}\hat{H}\ket{\psi(t)}$, containing a subextensive but large number of states, $\mathcal{N}\gg 1$.

Our model Hamiltonian \eqref{eq:TFIM} has no spectral degeneracies for finite $N$, and it is chaotic for small but non-zero values of $\alpha$ \cite{Russomanno2021}. This seems to imply that the standard ETH should suffice to describe thermal equilibrium states. However, according to this theory, the system must be always in a paramagnetic state where the order parameter vanishes, $\overline{\langle \hat{m}\rangle}=0$. This is because $\hat{m}$ changes the parity of any vector, and therefore $\bra{E_{n,\pm}}\hat{m}\ket{E_{n,\pm}}=0$. This is in stark contrast with the fact that Eq. \eqref{eq:TFIM} can display ferromagnetic phases with $\langle \hat{m}\rangle \neq 0$. In order to understand why, we use the framework developed in \cite{Gomez2025}, where a generalization of the ETH to account for symmetry-breaking has been developed. Since the mean level spacing of this model decreases \textit{exponentially} with $N$, it is always possible to find pairs of opposite-parity eigenstates,  $\{\ket{E_{n,+}},\ket{E_{n,-}}\}$, in which the quantum coherence survives during huge times, even for relatively small system sizes. These eigenspaces are built with states of opposite parity $E_{n,+}$, $E_{n,-}$, with minimum gap, $|E_{n,+}-E_{n,-}|$. As a consequence, any state belonging to the subspace $\mathcal{E}_n$ spanned by $\{\ket{E_{n,+}},\ket{E_{n,-}}\}$ behaves like an eigenstate of the Hamiltonian with energy $E_n \equiv E_{n,+}=E_{n,-}$ for all practical purposes. See Appendix \ref{sec:subspaces} for a discussion. The generalization of the ETH proposed in \cite{Gomez2025} relies on this idea. Applied to a system with a discrete $\mathbb{Z}_2$ symmetry, like the transverse-field Ising model, it works with the two dimensional subspaces $\mathcal{E}_n$. Within this framework, the key element to determine whether symmetry-breaking equilibrium states are possible or not is the projection of the order parameter, in this case $\hat{m}$, onto these subspaces; and, in particular, its eigenvalues, $\lambda_{n,+}=-\lambda_{n,-} \equiv \lambda_n$. In \cite{Gomez2025} it is proved that there exists ferromagnetic equilibrium states in the thermodynamic limit if and only if there exist eigenspaces where $\lim_{N \rightarrow \infty} |\lambda_n| \neq 0$. Therefore, the presence of values $|\lambda_n|$ significantly larger than zero in a particular spectral region is a strong signature of the existence of symmetry-breaking equilibrium states for any system size. 

\textit{Ferromagnetic scars.--} Fig. \ref{fig:phases} depicts the results for $\lambda_n$ in each of the individual subspaces $\mathcal{E}_n$, $J=1$, $h=0.1$ and $N=22$, and different values of $\alpha$. Figure \ref{fig:phases}(a) depicts the case of $\alpha=0.6$, corresponding to the strong long-range regime \cite{Defenu2023}. In the low energy region, all eigenvalues are nonzero. This generates a low-energy ferromagnetic phase with features that are similar to the fully-connected case $\alpha=0$, perhaps due to remnants  of stable scars arising from the conservation of $(\sum_{i}\pmb{\hat{\sigma}}_{i})^{2}$, as shown in \cite{Lerose2025}. We also find a mixed region, with zero and nonzero eigenvalues, that is quite wide in energy. This is compatible with the behavior of the model in the fully-connected limit $\alpha=0$ (see, e.g., Ref. \cite{Corps2024}). In the high-lying region of the spectrum, we find a paramagnetic spectral phase where the eigenvalues are zero. This region is quite small because $h \ll J$, a regime in which confinement is expected \cite{Vovrosh2022}. In Appendix \ref{sec:ferroapp} we illustrate how the size of this region depends on the value of $h/J$, vanishing when it is larger than the critical value of the quantum phase transition.

The general picture remains unchanged for values of $\alpha$ a little higher. For example, Fig. \ref{fig:phases}(b) displays the case of $\alpha=1.4$. This figure shows a behavior similar to that of $\alpha=0.6$, except that the mixed region becomes even wider in energy and the ferromagnetic region at low energies becomes less pronounced. The case of $\alpha=2.2$, depicted in Fig. \ref{fig:phases}(c), is much more interesting. Due to the absence of a ferromagnetic phase at $T>0$ for $\alpha>2$, we would expect a picture completely different for this case ---that all the values of $\lambda_n$ should be close to zero throughout the spectrum. Surprisingly, our results reveal the formation of bands in the spectrum whereby some values of $\lambda_n$ are close to zero, but some others are significantly different from zero. This band structure can be observed up to high excitation energies. We call these states \textit{ferromagnetic scars}, and they violate the generalized ETH. Indeed, these results suggest that initial conditions consisting in linear combinations of states in these bands with nonzero magnetization, will evolve towards a ferromagnetic equilibrium state, even though all thermal states are paramagnetic for $\alpha=2.2$. As shown in Figs. \ref{fig:phases}(d)-(f), the relative relevance of the ferromagnetic scars diminishes as $\alpha$ is further increased into the short-range interacting limit $\alpha\to\infty$, as the values of $\lambda_n$ increasingly cluster around the zero line. This means that ferromagnetic equilibrium states become less prevalent for high values of $\alpha$. To delve into this point, we have represented in Fig. \ref{fig:phases}(g) the scaling of the number of scarred states with system size for different values of the power-law exponent $\alpha$. This is computed as the number of subspaces $\mathcal{E}_n$ with $\lambda_n>0.2$ (other choices of this threshold are possible with qualitatively analogous results). The behavior for $\alpha=0.6,1.4$ and $2.2$ is very similar; the number of scarred states increases exponentially with $N$. This shows, in particular, that the scarred ferromagnetic states for $\alpha=2.2>\alpha_{c}=2$ play a significant role even when the model displays no thermal phase transition. For $\alpha=3$ we also observe an exponential growth with $N$, although the increase is slower than for the previous cases. Finally, the cases $\alpha=5,10$ do not seem to support an exponential growth with $N$.

\begin{center}
\begin{figure}[h!]
\hspace*{-0.84cm}\includegraphics[width=0.50\textwidth]{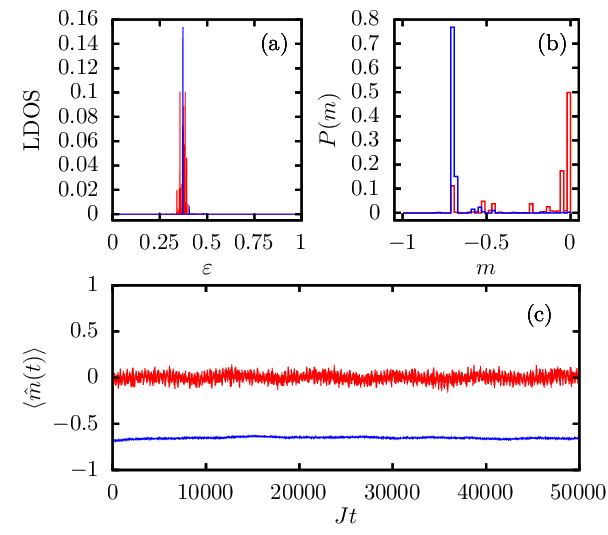}
\caption{(a) Local density of states for two initial states populating different regions of the energy spectrum. The initial states considered are $\ket{\psi_{1}}=\ket{\uparrow\downarrow\downarrow\downarrow\downarrow\downarrow\downarrow\uparrow\downarrow\downarrow\downarrow\downarrow\downarrow\downarrow\uparrow\downarrow\downarrow\downarrow\downarrow\downarrow}_{x}$ (blue) and $\ket{\psi_{2}}=\ket{\uparrow\uparrow\uparrow\uparrow\downarrow\downarrow\downarrow\downarrow\downarrow\downarrow\downarrow\uparrow\uparrow\uparrow\downarrow\downarrow\downarrow\downarrow\downarrow\downarrow}_{x}$ (red). (b) Distribution of the eigenvalues of the reduced total magnetization in each $\mathcal{E}_n$ subspace. (c) Time evolution of the normalized total magnetization for these two initial states. System size is $N=20$, and $\alpha=2.2$, $J=1$, $h=0.1$. }
\label{fig:ldos}
\end{figure}
\end{center}

A first conclusion emerging from these results is that there exists a large number of scarred ferromagnetic states for $\alpha \gtrsim 2$, where there is no ferromagnetic phase at $T>0$. But this is not enough to claim that there must be some observable consequences of this fact. Thus, our next step is to answer the following question: are these scarred states physically relevant? That is, can they be selectively populated by means of simple, experimentally accesible, protocols?  We tackle this question in Fig. \ref{fig:ldos}. Here, we study the dynamics of two initial states with similar average energy.  Both of them are initially magnetized along the x-axis; we label them as $\ket{\psi_{1}}=\ket{\uparrow\downarrow\downarrow\downarrow\downarrow\downarrow\downarrow\uparrow\downarrow\downarrow\downarrow\downarrow\downarrow\downarrow\uparrow\downarrow\downarrow\downarrow\downarrow\downarrow}_{x}$ and $\ket{\psi_{2}}=\ket{\uparrow\uparrow\uparrow\uparrow\downarrow\downarrow\downarrow\downarrow\downarrow\downarrow\downarrow\uparrow\uparrow\uparrow\downarrow\downarrow\downarrow\downarrow\downarrow\downarrow}_{x}$. In Fig. \ref{fig:ldos}(a) we represent the corresponding local density of states for each of these states as a function of $\varepsilon=(E-E_{\textrm{gs}})/(E_{\max}-E_{\textrm{gs}})$, where $E_{\textrm{gs}}$ is the ground-state energy, and $E_{\max}$ is the energy of the highest-excited state. Most of the population is concentrated around the $32\%$ central part of the spectrum for both initial states, although the values of the population coefficients are different; this corresponds with an inverse temperature $\beta=(k_B T)^{-1} \approx 0.46$. In Fig. \ref{fig:ldos}(b) we plot the distribution of the eigenvalues of the total magnetization for these two initial states. Here, $m$ denotes the eigenvalues of the order parameter $\hat{m}$. We clearly observe that $\ket{\psi_{1}}$ mainly populates non-zero values of the magnetization, while for $\ket{\psi_{2}}$ the distribution is mostly centered around $m\approx 0$. The ferromagnetic-paramagnetic nature of these states is confirmed in Fig. \ref{fig:ldos}(c), where we plot the long-time instantaneous value of the magnetization for $\ket{\psi_{1}}$ and $\ket{\psi_{2}}$. The equilibrium value for $\ket{\psi_{1}}$ is non-zero, $\overline{\hat{m}}_{\psi_{1}}\neq 0$, while it is close to zero for $\ket{\psi_{2}}$, $\overline{\hat{m}}_{\psi_{2}}\approx 0$.  The results of this figure illustrate the following. While the initial states $\ket{\psi_{1}}$ and $\ket{\psi_{2}}$ are similar in terms of energy, $\ket{\psi_{1}}$ mainly populates scarred ferromagnetic states and, as a consequence, the corresponding equilibrium state is ferromagnetic. On the contrary, the contribution of scarred ferromagnetic states is not significant for $\ket{\psi_{2}}$, and the ensuing equilibrium state is paramagnetic. As both of them are simple experimentally accessible states, like the ones used to study domain-wall confinement in \cite{Tan2021}, we can safely conclude that the ferromagnetic scarred states shown in Fig. \ref{fig:phases} play a very relevant role in the dynamics of the long-range transverse-field Ising model. 

The results shown in Fig. \ref{fig:ldos} exemplify the role of scarred states as non-thermal equilibrium states. The initial states $\ket{\psi_{1}}$ and $\ket{\psi_{2}}$ have the same energy expectation value and very similar eigenstate distributions, but only $\ket{\psi_{1}}$ remains magnetized.  This would be impossible if these scars were qualitatively similar to finite-temperature magnetized equilibrium states: in such a case, {\em both} initial states should relax to magnetized equilibrium states.

\begin{center}
\begin{figure}[h!]
\hspace*{-0.97cm}\includegraphics[width=0.49\textwidth]{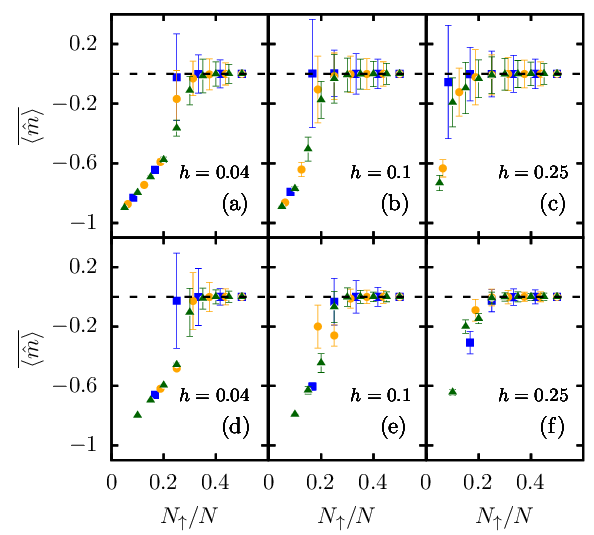}
\caption{Equilibrium value of the total magnetization for initial states with one ferromagnetic domain (top row) and two domains (bottom row), for different values of the magnetic field $h$, $\alpha=2.2$ and $J=1$. System size is $N=12,16,20$ (blue, orange, green). Errorbars represent the standard deviation with respect to the long-time average. Dashed horizontal lines indicate zero magnetization.}
\label{fig:mags}
\end{figure}
\end{center}

\textit{Ferromagnetic-paramagnetic transition.--} The main difference between the two initial states studied in Fig. \ref{fig:ldos} is that the first one is composed by three very small magnetic domains, whereas the second one is composed by two relatively large domains. Thus, it is natural to wonder if the number and size of such domains play the role of a control parameter, separating two different dynamical phases: one in which the ensuing equilibrium state is ferromagnetic, and another one in which the system evolves onto the expected paramagnetic equilibrium state.
We address this issue in Fig. \ref{fig:mags}. We study the equilibrium value of the total magnetization, $\overline{\langle \hat{m}\rangle}$, for different initial states composed of one (top row) or two (bottom row) magnetic domains, $\alpha=2.2$ and different values of $h$. These long-time averages are represented as a function of the ratio of up spins, $N_{\uparrow}/N$, and they have been calculated by averaging over the same time interval as results in Fig. \ref{fig:ldos}. The results for $h=0.04$ are fully compatible with the existence of two dynamical phases. For $N_{\uparrow}/N \lesssim 0.2$, we observe clear ferromagnetic equilibrium states, whereas for $N_{\uparrow}/N \gtrsim 0.2$ all the initial conditions relax to the expected paramagnetic equilibrium state. Note that all the initial states up to $N_{\uparrow}/N=1/2$ have negative magnetization, so those with $N_{\uparrow}/N \gtrsim 0.2$  lose their ferromagnetic nature as they approach the final equilibrium value, whereas those with $N_{\uparrow}/N \lesssim 0.2$ do not. For $h=0.1$, we observe a very similar scenario, with a smaller critical value for $N_{\uparrow}/N$; in this case, the results for initial conditions with just one magnetic domain are clearer than the ones in which the initial conditions have two magnetic domains. The case with $h=0.25$ is much less clear, although the results obtained with the largest system ($N=20$) suggest that there also exist a ferromagnetic scarred phase at low values of $N_{\uparrow}/N$. 

From all these results, we propose that the long-range transverse-field Ising model with small values of $h/J$ and $\alpha \gtrsim 2$ has a {\em scarred ferromagnetic phase}. Initial states with low values of $N_{\uparrow}/N$ and a small number of magnetic domains mainly populate scarred ferromagnetic scars, and therefore they evolve onto ferromagnetic equilibrium state. On the other hand, initial conditions with more and/or larger magnetic domains, or no magnetic structure at all, mainly populate non-scarred states and thus they evolve towards the canonical paramagnetic equilibrium value. We conjecture that a critical value of $N_{\uparrow}/N$, probably depending on $h/J$, $\alpha$ and the number of magnetic domains in the initial condition, separates the scarred ferromagnetic and the normal phases.

\textit{Discussion.--} Our results can explain a number of previous findings concerning equilibration in the TFIM. In Ref. \cite{Liu2019} it is shown that a nonzero value of the magnetization can be preserved up to very long time scales even for values of $\alpha$ large enough to exclude a thermal ferromagnetic phase. To explain this finding, the authors proposed a model where low-lying excitations behave as spin waves consisting in a single magnetic domain. Although consistent with the theory in \cite{Liu2019}, the results presented here can explain stable ferromagnetism for initial states with an arbitrary number of domains, as we have exemplified with two and three domains. A similar idea is discussed in \cite{Collura2022}, where it is shown that the magnetization remains different from zero in a kicked TFIM with $\alpha>2$ due to confinement. Our results suggest that these long-standing ferromagnetic states are real equilibrium ones; they are induced by ferromagnetic scars, and their survival time is expected to grow exponentially with system size, therefore becoming larger than any observational time for a quite small number of particles. In Refs. \cite{Homrighausen2017,Halimeh2017,Zunkovic2018} the authors study dynamical phase transitions and ferromagnetic phases as a function of $\alpha$. A conclusion of \cite{Homrighausen2017,Halimeh2017} is that ferromagnetic states can be found for almost every $\alpha$, whereas in \cite{Zunkovic2018} this can only be guaranteed up to $\alpha=2$, but with some inconclusive results for values of $\alpha$ slightly larger than $2$, especially for small values of $h/J$. Our results suggest that the ferromagnetic phase exists for a certain range of $\alpha>2$, due to the exponential growth of the number of ferromagnetic scars with $N$, shown in Fig. \ref{fig:phases}, and the dynamical results illustrated in Figs. \ref{fig:ldos} and \ref{fig:mags}. We expect that it should disappear for sufficiently large values of $\alpha$ and $h/J$. These two results seem to be compatible. Finally, the existence of ferromagnetic scarred states may favor confinement in the chain. Symmetry-breaking equilibrium states in these kinds of models have been shown to be linked to the existence of a constant of motion \cite{Corps2021,Gomez2025,Corps2022,Corps2022prl,Corps2024,Corps2024pre}, restricting domain wall free propagation. Further research is needed to make a definite claim.

\textit{Acknowledgments.--}
A. L. C. acknowledges fruitful discussions with P. Cejnar, P. Stránský and J. Novotný. A.R. acknowledges financial support by the Spanish grant
PID2022-136285NB-C31 funded by Ministerio de Ciencia e Innovación/Agencia Estatal de Investigación MCIN/AEI/10.13039/501100011033 and FEDER “A Way of Making Europe”. A.L.C. acknowledges financial support from the Czech Science Foundation under project No. 25-16056S as well as the JUNIOR UK Fund project carried out at the Faculty of Mathematics and Physics, Charles University. The data that support the findings of this article are openly available \cite{data}.


\begin{appendix}
 
\section{Symmetries of the TFIM}\label{sec:simulationTFIM}

A simple way to account for all symmetries of the model is to consider that the TFIM with periodic boundary conditions is arranged in a closed ring with $N$ sites. Then, the Hamiltonian in Eq. \eqref{eq:TFIM} is invariant under all the transformations belonging to the dihedral $D_N$ group. As the generators of this group are a $2 \pi/N$ rotation, $\hat{\mathcal{R}}$, and the inversion around any of symmetry axis of a regular polygon of $N$ sites, $\hat{\mathcal{I}}$, we build our computational basis taking into account these two transformations:

(1) Invariance under inversion symmetry, $\hat{\mathcal{I}}$. We choose the symmetry axis passing through the center of the chain. In the usual site basis with states $\ket{\phi}=\bigotimes_{i=1}^{N}\ket{\phi_{i}}_{z}$, $\phi_{i}\in\{\uparrow,\downarrow\}$, this operator induces a reflection along the center of the chain, $\hatmath{I}\ket{\phi_{1}\,\phi_{2}\,\ldots\,\phi_{N}}_{z}=\ket{\phi_{N}\,\ldots\,\phi_{2}\,\phi_{1}}_{z}$. Our computational basis consists of states $\ket{\phi}$ in the positive inversion sector, fulfilling $\hatmath{I}\ket{\phi}=\ket{\phi}$. 

(2) $2 \pi/N$ rotation symmetry, $\hat{\mathcal{R}}$. This operator acts on the site basis as $\hatmath{R}\ket{\phi_{1}\,\phi_{2}\,\cdots,\phi_{N}}_{z}=\ket{\phi_{N}\,\phi_{1}\,\cdots,\phi_{N-1}}_{z}$. 

Note that $\hat{\mathcal{R}}$ and $\hat{\mathcal{T}}$ do not commute in general, but they do in the zero and $\pi$ momentum sectors. We take advantage of this fact by building our computational basis by means of states $\ket{\phi}$ with zero momentum and positive inversion, fulfilling $\hatmath{I}\ket{\phi}=\ket{\phi}$, and $\hatmath{R}\ket{\phi}=\ket{\phi}$. Other sectors require more computational resources, and they give rise to the same qualitative results.

(3) Finally, Eq. \eqref{eq:TFIM} is invariant under the parity operator, $\hat{\Pi}=\prod_{i=1}^{N}\hat{\sigma}_{i}^{z}$ \cite{Russomanno2021}, commuting with both $\hat{\mathcal{I}}$ and $\hat{\mathcal{T}}$. Our basis contains states belonging to both parity sectors.

Another consequence of the use of PBCs is that the $\alpha$-dependent interaction potential in Eq. \eqref{eq:TFIM} is computed as $\frac{J}{\mathcal{K}(\alpha)}D_{ij}^{-\alpha}$, where $D_{ij}=\min\{|i-j|,N-|i-j|\}$ with $\mathcal{K}(\alpha)=\frac{1}{N-1}\sum_{i < j}D_{ij}^{-\alpha}$ being the Kac factor \cite{Kac1963}, which ensures the Hamiltonian intensiveness when $\alpha<1$ but could be omitted for $\alpha\geq 1$. 

\section{Construction of eigenstate subspaces}\label{sec:subspaces}

\begin{center}
\begin{figure}[h!]
\hspace*{-0.8cm}\includegraphics[width=0.49\textwidth]{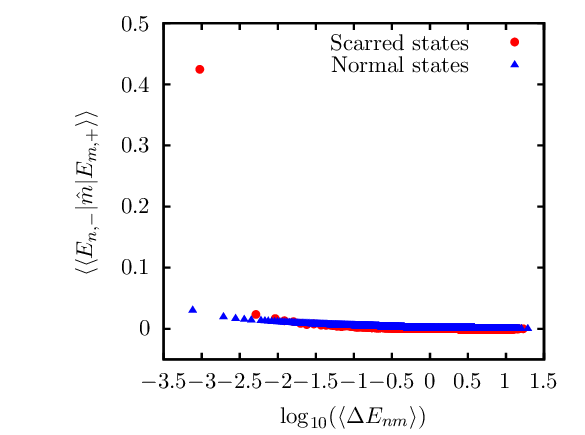}
\caption{Absolute expectation value of the intensive magnetization, $\hat{m}=2\hat{M}/N$, over the Hamiltonian eigenstates of opposite parity, $\bra{E_{n,-}}\hat{m}\ket{E_{n,+}}$, as a function of the gap $\Delta E_{nm}=|E_{n,+}-E_{m,-}|$. Results are averages over all possible combinations of Hamiltonian eigenstates. Red points correspond to expectation values in the case of scarred states, and red triangles correspond to normal states (see text). System parameters are $J=1$, $h=0.1$, $\alpha=2.2$ and $N=20$.}
\label{fig:gapeigenvalues}
\end{figure}
\end{center}

In the main text we consider the energy subspaces $\mathcal{E}_{n}$ composed of eigenstates of opposite parity, and we calculate the eigenvalues, $\lambda$, of the magnetization restricted to these $2$-dimensional subspaces. To determine the pairs of eigenstates in such doublets, we fix a state in a parity sector and consider eigenvalues in the opposite parity sector such that the gap $\Delta E_{nm}=|E_{n,+}-E_{m,-}|$ approaches zero in the TL. In realistic physical systems, such as ours, the Hilbert space dimension grows exponentially with $N$, but the spectral width grows linearly with $N$. Therefore, we expect the number of pairs of eigenvalues that satisfy this condition to be quite large \cite{Gomez2025}. To take care of this ambiguity, to construct the $2$-dimensional energy subspaces we consider the states $\ket{E_{n,+}}$ and $\ket{E_{m,-}}$ such that the gap $\Delta E_{nm}$ is minimal when $E_{n,+}$ is fixed and all $E_{m,-}$ are taken into account. Here we show that this criterion works very well and can be used to construct such eigenspaces. 

In Ref. \cite{Corps2021}, and subsequently in Refs. \cite{Corps2022,Corps2022prl}, we showed that breaking of a $\mathbb{Z}_2$ symmetry is linked to the appearance of a discrete constant of motion, $\hat{\mathcal{C}}$, whose eigenstates consist of linear combinations of two Hamiltonian eigenstates of opposite parity. A consequence of this theory is that the non-diagonal matrix element of an intensive version of the order parameter within this subspace, $|\bra{E_{m,-}} \hat{m} \ket{E_{n,+}}|$, is significantly different from zero, whereas the corresponding matrix elements in other subspaces, $|\bra{E_{k,-}} \hat{m} \ket{E_{n,+}}|$, with $k \neq m$ are typically very close to zero. In other words, for a given $E_{n,+}$ there is a privileged partner, $E_{m,-}$, not necessarily with $n=m$, that can be used to build the eigenspaces $\mathcal{E}_{n}$. This is corroborated in Fig. \ref{fig:gapeigenvalues}, where we show the average eigenstate expectation value of $\hat{m}$ over all possible combinations of Hamiltonian eigenstates of opposite parity as a function of the average gap for a case with $\alpha=2.2$ and $N=20$ spins. We separate results for scarred and normal states. We have considered that a state is scarred if the corresponding $\hat{m}$ eigenvalue satisfies $\lambda>0.3K$ with $K\equiv\max_{i}\lambda_{i}$, and it is normal otherwise. For scarred states, the off-diagonal elements outside the energy subspace are much smaller than the eigenvalues. By constrast, for non-scarred states this difference is not so clear, which shows that such eigenspaces are not as relevant in this case.

\section{Ferromagnetic scars for other parameter values}\label{sec:ferroapp}

\begin{center}
\begin{figure}[t!]
\hspace*{-0.8cm}\includegraphics[width=0.51\textwidth]{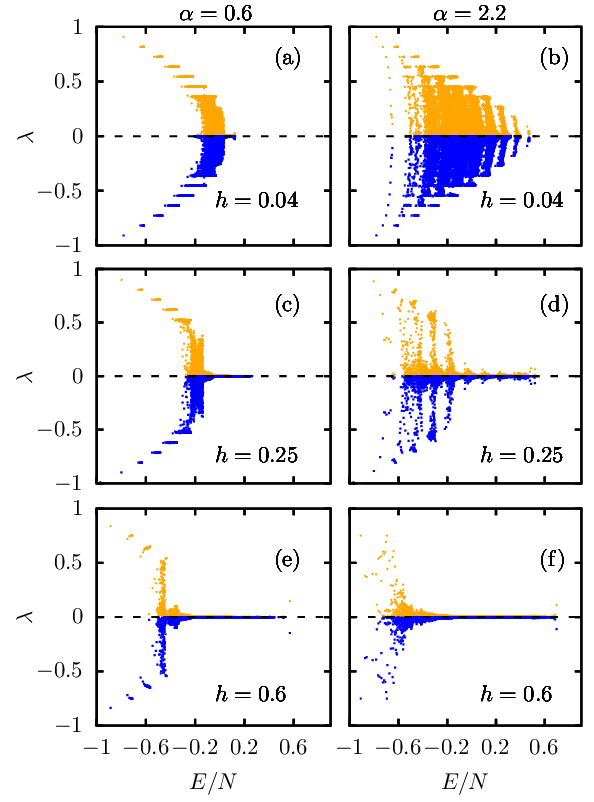}
\caption{(a)-(f) Eigenvalues of the normalized total magnetization in individual energy eigenspaces for different values of the power-law interaction, $\alpha$, $h$, and $J=1$ as a function of eigenspace energy. System size is $N=22$.}
\label{fig:phasesapp}
\end{figure}
\end{center}

In Fig. \ref{fig:phasesapp} we report results analogous to those in Fig. \ref{fig:phases} in the main text, but for two values of the power-law exponent $\alpha=0.6,2.2$ and for three values of the magnetic field, $h=0.04,0.25,0.6$. For very long-range interactions, $\alpha=0.6$, the size of the ferromagnetic phase decreases as $h$ increases, leading to an increasingly large paramagnetic region. For $\alpha=2.2$, where the scarred ferromagnetic phase was identified in Fig. \ref{fig:phases}, we observe that increasing $h$ also has the effect of destroying these scarred states, leading to a standard paramagnetic phase for sufficiently large values of $h$.

\end{appendix}


\begin{thebibliography}{100}

\bibitem{Defenu2023} N. Defenu, T. Donner, T. Marcrì, G. Pagano, S. Ruffo, and A. Trombettoni, \textit{Long-range interacting quantum systems}, Rev. Mod. Phys. \textbf{95}, 035002 (2023).

\bibitem{Defenu2024} N. Defenu, A. Leores, and S. Pappalandi, \textit{Out-of-equilibrium dynamics of quantum many-body systems with long-range interactions}. Phys. Rep. \textbf{1074}, 1 (2024).

\bibitem{Dyson1969} F. J. Dyson, \textit{Existence of a phase transition in an one-dimensional Ising ferromagnet}, Commun. Math. Phys. \textbf{12}, 91 (1969).

\bibitem{Thouless1969} D. J. Thouless, \textit{Long-range order in one-dimensional Ising systems}, Phys. Rev. \textbf{187}, 732 (1969).

\bibitem{Fisher1972} M. E. Fisher, S.-k. Ma, and B. G. Nickel, \textit{Critical exponents for long-range interactions}, Phys. Rev. Lett. \textbf{29}, 917 (1972).

\bibitem{Liujten1997} E. Luijten and H. W. J. Blo\"ote, \textit{Classical critical behavior of spin models with long-range interctions}, Phys. Rev. B \textbf{56}, 8945 (1997).

\bibitem{Gonzalez-Lazo2021} E. Gonzalez-Lazo, M. Heyl, M. Dalmonte, and A. Angelone, \textit{Finite-temperature critical behavior of long-range quantum Ising models}, SciPost Phys. \textbf{11}, 076 (2021).

\bibitem{Corps2024pre} A. L. Corps and A. Rela\~{n}o, General theory for discrete symmetry-breaking equilibrium states in quantum systems, Phys. Rev. E \textbf{110}, 034137 (2024).

\bibitem{Schuckert2025} A. Schuckert, O. Katz, L. Feng, E. Crane, A. De, M. Hafezi, A. V. Gorshkov, and C. Monroe, \textit{Observation of a finite-energy phase transition in a one-dimensional quantum simulator}, Nat. Phys. \textbf{21}, 374 (2025).

\bibitem{Landau1980} L. Landau and E. Lifshitz, \textit{Statistical Physics} (Butterworth-Heinemann, 1980).

\bibitem{Zhang2017} J. Zhang, P. W. Hess, A. Kyprianidis, P. Becker, A. Lee, J. Smith, G. Pagano, I.-D. Potirniche, A. C. Potter, A. Vishwanath, N. Y. Yao, and C. Monroe, \textit{Observation of a discrete time crystal}, Nature \textbf{543}, 217 (2017).

\bibitem{Choi2017} S. Choi, J. Choi, R. Landig, G. Kucsko, H. Zhou, J. Isoya, F. Jelezko, S. Onoda, H. Sumiya, V. Khemani, C. von Keyserlingk, N. Y. Yao, E. Demler, and M. D. Lukin, \textit{Observation of discrete time-crystalline order in a disordered dipolar many-body system}, Nature \textbf{543}, 221 (2017).

\bibitem{Collura2022} M. Collura, A. De Luca, D. Rossini, and A. Lerose, \textit{Discrete Time-Crystalline Response Stabilized by Domain-Wall Confinement}, Phys. Rev. X \textbf{12}, 031037 (2022).

\bibitem{Gargiulo2024} R. Gargiulo, G. Passarelli, P. Lucignano and A. Russomanno, \textit{Swapping Floquet time crystal}, Phys. Rev. B \textbf{109}, 174310 (2024).

\bibitem{Pizzi2021} A. Pizzi, J. Knolle and A. Nunnenkamp, \textit{Higher-order and fractional discrete time crystals in clean long-range interacting systems}, Nat. Commun. \textbf{12}, 2341 (2021).

\bibitem{Yu2019} W. C. Yu, J. Tangpanitanon, A. W. Glaetzle, D. Jaksch and D. G. Angelakis, \textit{Discrete time crystal in globally driven interacting quantum systems without disorder}, Phys. Rev. A \textbf{99}, 033618.

\bibitem{Yu2025} H. Yu and T.-C. Wei, \textit{Robustness condition for general disordered discrete time crystals: Subspace-thermal discrete time crystals from phase transitions between different n-tuple discrete time crystals}, Phys. Rev. B \textbf{111}, 174311 (2025). 

\bibitem{Russomanno2021} A. Russomanno, M. Fava and M. Heyl, Quantum chaos and ensemble inequivalence of quantum long-range Ising chains, Phys. Rev. B \textbf{104}, 094309 (2021).

\bibitem{Halimeh2017b} J. C. Halimeh, V. Zauner-Stauber, I. P. McCulloch, I. de Vega, U. Schollw\"ock, and M. Kastner, \textit{Prethermalization and persisent order in the absence of a thermal phase transition}, Phys. Rev. B \textbf{95}, 024302 (2017).

\bibitem{Neyenhuis2017} B. Neyenhuis, J. Zhang, P. W. Heiss, J. Smith, A. C. Lee, P. Richerme, Z.-X. Gong, A. V. Gorshkov, and C. Monroe, \textit{Observation of prethermalization in long-range interacting spin chains}, Sci. Adv. \textbf{3}, e1700672 (2017).

\bibitem{Defenu2021} N. Defenu, \textit{Metastability and discrete spectrum of long-range systems}, Proc. Natl. Acad. Sci. U.S.A. \textbf{118}, e2101785118 (2021).

\bibitem{Mattes2025} R. Mattes, I. Lesanovsky, and F. Carollo, \textit{Long-Range Interacting Systems Are Locally Noninteracting}, Phys. Rev. Lett. \textbf{134}, 070402 (2025).

\bibitem{Zhang2017b} J. Zhang, G. Pagano, P. W. Hess, A. Kyprianidis, P. Becker, H. Kaplan, A. V. Gorshkov, Z.-X. Gong, and C. Monroe, \textit{Observation of a many-body dynamical phase transition with a 53-qubit quantum simulator}, Nature \textbf{551}, 601 (2017).

\bibitem{Halimeh2017} J. C. Halimeh and V. Zauner-Stauber, \textit{Dynamical phase diagram of quantum spin chains with long-range interactions}, Phys. Rev. B \textbf{96}, 134427 (2017).

\bibitem{Homrighausen2017} I. Homrighausen, N. O. Abeling, V. Zauner-Stauber, and J. C. Halimeh, \textit{Anomalous dynamical phase in quantum spin chains with long-range interactions}, Phys. Rev. B \textbf{96}, 104436 (2017).

\bibitem{Zunkovic2018} B. Žunkovič, M. Heyl, M. Knap and A. Silva, \textit{Dynamical Quantum Phase Transitions in Spin Chains with Long-Range Interactions: Merging Different Concepts of Nonequilibrium Criticality}, Phys. Rev. Lett. \textbf{120}, 130601.

\bibitem{Piccitto2019} G. Piccitto, B. Zunkovic, and A. Silva, \textit{Dynamical phase diagram of a quantum Ising chain with long-range interactions}, Phys. Rev. B \textbf{100}, 180402(R) (2019).

\bibitem{Ranabhat2022} N. Rananhat and M. Collura, \textit{Dynamics of the order parameter statistics in the long range Ising model}, SciPost Phys. \textbf{12}, 126 (2022).

\bibitem{Corps2022} A. L. Corps and A. Rela\~{n}o, \textit{Dynamical and excited-state quantum phase transitions in collective systems}, Phys. Rev. B \textbf{106}, 024311 (2022).

\bibitem{Corps2022prl} A. L. Corps and A. Rela\~{n}o, \textit{Theory of Dynamical Phase Transitions in Quantum Systems with Symmetry-Breaking Eigenstates}, Phys. Rev. Lett. \textbf{130}, 100402 (2023).

\bibitem{Joshi2022} M. K. Joshi, F. Kranzl, A. Schuckert, I. Lovas, C. Maier, R. Blatt, M. Knap, and C. F. Roos, \textit{Observing emergent hydrodynamics in a long-range quantum magnet}, Science \textbf{376}, 720 (2022).

\bibitem{Hauke2013} R. Hauke and L. Tagliacozzo, \textit{Spread of Correlations in Long-Range Interacting Quantum Systems}, Phys. Rev. Lett. \textbf{111}, 207202 (2013).

\bibitem{Lerose2019} A. Lerosse, B. Zunkovic, A. Silva, and A. Gambassi, \textit{Quasilocalized excitations induced by long-range interactions in translationally invariant quantum spin chains}, Phys. Rev. B \textbf{99}, 121112(R) (2019).

\bibitem{Liu2019} F. Liu, R. Lundgren, P. Titum, G. Pagano, J. Zhang, C. Monroe and A. V. Gorshkov, \textit{Confined Quasiparticle Dynamics in Long-Range Interacting Quantum Spin Chains}, Phys. Rev. Lett. 122, 150601 (2019).

\bibitem{Tan2021} W. L. Tan, P. Becker, F. Liu, G. Pagano, K. S. Collins, A. De, L. Feng, H. B. Kaplan, A. Kyprianidis, R. Lundgren, W. Morong, S. Whitstt. A. V. Gorshkov, and C. Monroe, \textit{Domain-wall confinement and dynamics in a quantum simulator}, Nat. Phys. \textbf{17}, 742 (2021).

\bibitem{Verdel2020} R. Verdel, F. Liu, S. Whitsitt, A. V. Gorshkov, and M. Heyl, \textit{Real-time dynamics of string breaking in quantum spin chains}, Phys. Rev. B \textbf{102}, 014308 (2020).

\bibitem{Vovrosh2022} J. Vovrosh, R. Mukherjee, A. Bastianello, and J. Knolle, \textit{Dynamical Hadron Formation in Long-Range Interacting Quantum Spin Chains}, PRX Quantum 3, 040309 (2022).

\bibitem{Mivehvar2021} F. Mivehvar, F. Piazza, T. Donner, and H. Ritsch, \textit{Cavity
QED with quantum gases: New paradigms in many-body physics}, Adv. Phys. \textbf{70}, 1 (2021).

\bibitem{Britton2012} J. W. Britton, B. C. Sawyer, A. C. Keith, C.-C. J. Wang, J. K.
Freericks, H. Uys, M. J. Biercuk, and J. J. Bollinger, \textit{Engineered two-dimensional Ising interactions in a
trapped-ion quantum simulator with hundreds of spins}, Nature (London) \textbf{484}, 489 (2012).

\bibitem{Feng2023} L. Feng, O. Katz, C. Haack, M. Maghrebi, A. V. Gorshkov, Z. Gong, M. Cetina, and C. Monroe, \textit{Continuous symmetry breaking in a trapped-ion spin chain}, Nature (London) \textbf{623}, 713 (2023).

\bibitem{Labuhn2016} H. Labuhn, D. Barredo, S. Ravets, S. de L\'{e}s\'{e}leuc, T. Macrì,
T. Lahaye, and A. Browaeys, \textit{Tunable two-dimensional arrays of single Rydberg atoms for realizing quantum Ising models}, Nature (London) \textbf{534}, 667 (2016).

\bibitem{Zeiher2017} J. Zeiher, J. Y. Choi, A. Rubio-Abadal, T. Pohl, R. van Bijnen, I. Bloch, and C. Gross, \textit{Coherent many-body spin dynamics in a long-range interacting Ising chain}, Phys. Rev. X \textbf{7}, 041063 (2017).

\bibitem{Chen2023} C. Chen, G. Bornet, M. Bintz, G. Emperauger, L. Leclerc, V. S. Liu, P. Scholl, D. Barredo, J. Hauschild, S. Chatterjee, M. Schuler, A. M. Läuchli, M. P. Zaletel, T. Lahaye, N. Y. Yao, and A. Browaeys, \textit{Continuous symmetry breaking in a two-dimensional Rydberg array}, Nature (London) \textbf{616}, 691 (2023).

\bibitem{Achutha2025} R. Achutha, D. Kim, Y. Kimura, and T. Kuwahara, \textit{Provably Efficient Simulation of 1D Long-Range Interacting Systems at Any Temperature}, Phys. Rev. Lett. \textbf{134},190404 (2025).

\bibitem{Turner2018} C. J. Turner, A. A. Michailidis, D. A. Abanin, M. Serbyn, and Z. Papi\'c, \textit{Weak ergodicity breaking from quantum many-body scars}, Nat. Phys. \textbf{14}, 745 (2018).

\bibitem{Lerose2025} A. Lerose, T. Parolini, R. Fazio, D. A. Abanin, and S. Pappalardi, \textit{Theory of Robust Quantum Many-Body Scars in Long-Range Interacting Systems}, Phys. Rev. X \textbf{15}, 011020 (2025).

\bibitem{Kac1963} M. Kac and E. Helfand, \textit{Study of several lattice systems with long range forces}, J. Math. Phys. (N.Y.) \textbf{4}, 1078 (1963).

\bibitem{Lipkin1965} H. Lipkin, N. Meshkov, and A. Glick, \textit{Validity of many-body approximation methods for a solvable model: (I). Exact solutions and perturbation theory}, Nucl. Phys. \textbf{62}, 188 (1965).

\bibitem{Botet1982} R. Botet, R. Jullien, and P. Pfeuty, \textit{Size scaling for infinitely coordinated systems}, Phys. Rev. Lett. \textbf{49}, 478 (1982).

\bibitem{Cejnar2021} P. Cejnar, , P. Str\'ansk\'y, M. Macek, and M. Kloc, \textit{Excited-state quantum phase transitions},  	J. Phys. A: Math. Theor. \textbf{54} (2021) 133001.

\bibitem{Puebla2015} R. Puebla, and A. Rela\~{n}o, \textit{Irreversible processes without energy dissipation in an isolated Lipkin-Meshkov-Glick model}, Phys. Rev. E \textbf{92}, 012101 (2015).

\bibitem{Puebla2013b} R. Puebla, and A. Rela\~{n}o, \textit{Non-thermal excited-state quantum phase transitions}, EPL \textbf{104}, 50007 (2013).

\bibitem{Sachdev} S. Sachdev, \textit{Quantum Phase Transitions}, (Cambridge University Press, 2011).

\bibitem{GonzalezLazo2021} E. Gonzalez Lazo, M. Heyl, M. Dalmonte, and A. Angelone, \textit{Finite-temperature critical behavior of long-range quantum Ising models}, SciPost Phys \textbf{11}, 076 (2021).

\bibitem{Sehrawat2021} A. Sehrawat, C. Srivastava, and U. Sen, \textit{Dynamical phase transitions in the fully connected quantum Ising model: Time period and critical time}, Phys. Rev. B \textbf{104}, 085105 (2021).

\bibitem{Halimeh2020} J. C. Halimeh, M. V. Damme, V. Zauner-Stauber and L. Vanderstraeten, \textit{Quasiparticle origin of dynamical quantum phase transitions}, Phys. Rev. Research \textbf{2}, 033111 (2020).

\bibitem{Deutsch1991} J. M. Deutsch, \textit{Quantum statistical mechanics in a closed system}, Phys. Rev. A \textbf{43}, 2046 (1991).

\bibitem{Srednicki1994} M. Srednicki, \textit{Chaos and quantum thermalization}, Phys. Rev. E \textbf{50}, 888 (1994).

\bibitem{Rigol2008} M. Rigol, V. Dunjko, and M. Olshanii, \textit{Thermalization and its mechanism for generic isolated quantum systems}, Nature \textbf{452}, 854 (2008).

\bibitem{Polkovnikov2011} A. Polkovnikov, K. Sengupta, A. Silva, and M. Vengalattore, \textit{Colloquium: Nonequilibrium dynamics of closed interacting quantum systems}, Rev. Mod. Phys. \textbf{83}, 863 (2011).

\bibitem{Gogolin2016} C. Gogolin and J. Eisert, \textit{Equilibration, thermalisation, and the emergence of statistical mechanics in closed quantum systems}, Rep. Prog. Phys. \textbf{79}, 056001 (2016).

\bibitem{DAlessio2016} L. D'Alessio, Y. Kafri, A. Polkovnikov, and M. Rigol, \textit{From quantum chaos and eigenstate thermalization to statistical mechanics and thermodynamics}, Adv. Phys, \textbf{65}, 239 (2016).

\bibitem{Gomez2025} S. Gómez, A. L. Corps and A. Rela\~{n}o, \textit{Quantum thermalization mechanism and the emergence of symmetry-breaking phases}, arXiv:2506.13370 [quant-ph].

\bibitem{Corps2024} A. L. Corps, A. Relaño and J. C. Halimeh, \textit{Unifying finite-temperature dynamical and excited-state quantum phase transitions}, Phys. Rev. Res \textbf{6}, 043080 (2024).

\bibitem{Corps2021} A. L. Corps and A. Rela\~{n}o, \textit{Constant of motion identifying excited-state quantum phases}, Phys. Rev. Lett. \textbf{127}, 130602 (2021).

\bibitem{data} Á. L. Corps and A. Rela\~{n}o, Data supporting Figs. 1-5 of manuscript at https://arxiv.org/abs/2507.16421. Zenodo, https://doi.org/10.5281/zenodo.18704812 (2026). 


\end{thebibliography}
\end{document}